# Metallic Hydrogen Sublattice and Proton Mobility in Copper Hydride at High Pressure

Thomas Meier[1]*, Florian Trybel[1], Giacomo Criniti[1], Dominique Laniel[2], Saiana Khandarkhaeva[1], Egor Koemets[1], Timofey Fedotenko[2], Konstantin Glazyrin[3], Michael Hanfland[4], Gerd Steinle-Neumann[1], Natalia Dubrovinskaia[2,5] and Leonid Dubrovinsky[1]

[1]Bayerisches Geoinstitut, Universität Bayreuth, 95440 Bayreuth, Germany
[2]Material Physics and Technology at Extreme Conditions, Laboratory of Crystallography, Universität Bayreuth, 95440 Bayreuth, Germany
[3]Deutsches Elektronen-Synchrotron (DESY), 22603, Hamburg, Germany
[4]European Synchrotron Radiation Facility (ESRF), 38043 Grenoble Cedex, France
[5]Department of Physics, Chemistry and Biology (IFM), Linköping University, SE-581 83, Linköping, Sweden
*thomas.meier@uni-bayreuth.de

Atomic and electronic structures of $Cu_2H$ and $CuH$ have been investigated by high pressure NMR spectroscopy, X-ray diffraction and *ab-initio* calculations. Metallic $Cu_2H$ was synthesized at a pressure of 40 GPa, and semi-metallic $CuH$ at 90 GPa, found stable up to 160 GPa. Experiments and computations suggest the formation of a metallic $^1H$-sublattice as well as a high $^1H$ mobility of $\sim 10^{-7}$ cm$^2$/s in $Cu_2H$. Comparison of $Cu_2H$ and $FeH$ data suggests that deviations from Fermi gas behavior, formation of conductive hydrogen networks, and high $^1H$ mobility could be common features of metal hydrides.

With their potential as hydrogen storage and high-temperature ($T$) superconducting materials [1], the interest in metal hydrides in condensed matter physics has been strong since their discovery [2] and they continue to attract significant research efforts [3,4]. Characterizing their stability, composition, electronic and transport properties provides crucial insight for applications, but metal hydrides continue to be enigmatic due to the highly variable metal-hydrogen bond. Following hydrogen uptake at elevated pressure ($P$) (and $T$), the electronic structure of metal hydrides can change significantly giving rise to their semiconducting [5], metallic [6] to even superconducting behavior [7].

Experimental probes in the diamond anvil cell (DAC), the most important device used to generate high-$P$ in solid state research of novel hydrides, have not been able to directly investigate electronic and dynamic properties of the hydrogen sub-systems until recently. By combining magnetic flux tailoring techniques [8–10] and modified DAC design using ion beam milling, the accessible $P$-range for $^1$H-NMR in the DAC has recently been extended tenfold from <10 GPa [11,12] to beyond 100 GPa [13,14].

While the formation of a metallic hydrogen sublattice in iron monohydride (FeH) has recently been described by our group [14] employing $^1$H-NMR in combination with *ab-initio* electronic structure calculations, no systematic study on the formation or dynamic properties of such sublattices have been conducted in comparable metal hydrides. Here, we present the results of investigations on copper hydrides using a combination of single crystal synchrotron X-ray diffraction, *in-situ* high-$P$ $^1$H-NMR, and density functional theory (DFT) based computations, which enabled establishing of their structural and electronic properties, as well as $^1$H mobility, and comparing them quantitatively to the results for FeH [14].

Hexagonal copper hydride (with wurtzite-type structure) was one of the first metal hydrides discovered in the XIX century [2]. As a highly reactive compound, it readily decomposes to the elements [15]. Over the past two decades, a number of high-$P$ copper hydrides were synthesized in the DAC, using dehydrogenated Cu and H$_2$, cubic $CuH_{0.4}$ with an *fcc* arrangement of Cu atoms was formed in the DAC at $P$>14 GPa [16]. By contrast, in the same work no reaction was observed when Cu foil was used as starting materials, suggesting that $CuH_{0.4}$ does not form at equilibrium, rather, Burtovyy and Tkacz [16] inferred that the presence of small crystallites from the dehydrogenated copper hydride seems required for nucleation. At $P$>18 GPa and room $T$, a trigonal $Cu_2H$ with the anti-CdI$_2$ structure was synthesized and found stable to 50 GPa [17]. Donnerer *et al.* [17] also observed an *fcc*-based copper hydride after decompression of trigonal $Cu_2H$ and recompression to 12.5 GPa, with a reported composition of $CuH_{0.15}$, a significantly lower degree of hydrogenation. Laser heating of pure copper grains with excess of hydrogen resulted in the formation of $Cu_2H$ at ~30 GPa, and $CuH_{0.65}$ at ~50 GPa [18], which is cubic with an *fcc* arrangement of Cu atoms. To this day, $CuH_{0.65}$ is the group 11 metal hydride with the highest documented hydrogen content at high $P$.

In this work, all $CuH_x$ compounds have been synthesized by direct reaction of pure copper and paraffin in the laser-heated DAC as described in our previous work [14,19]. The preparation procedure for all DAC experiments, $P$-determination, sample synthesis, details of the XRD studies, the NMR experiments, and the DFT-based calculations are outlined in the Supplemental Material [20].

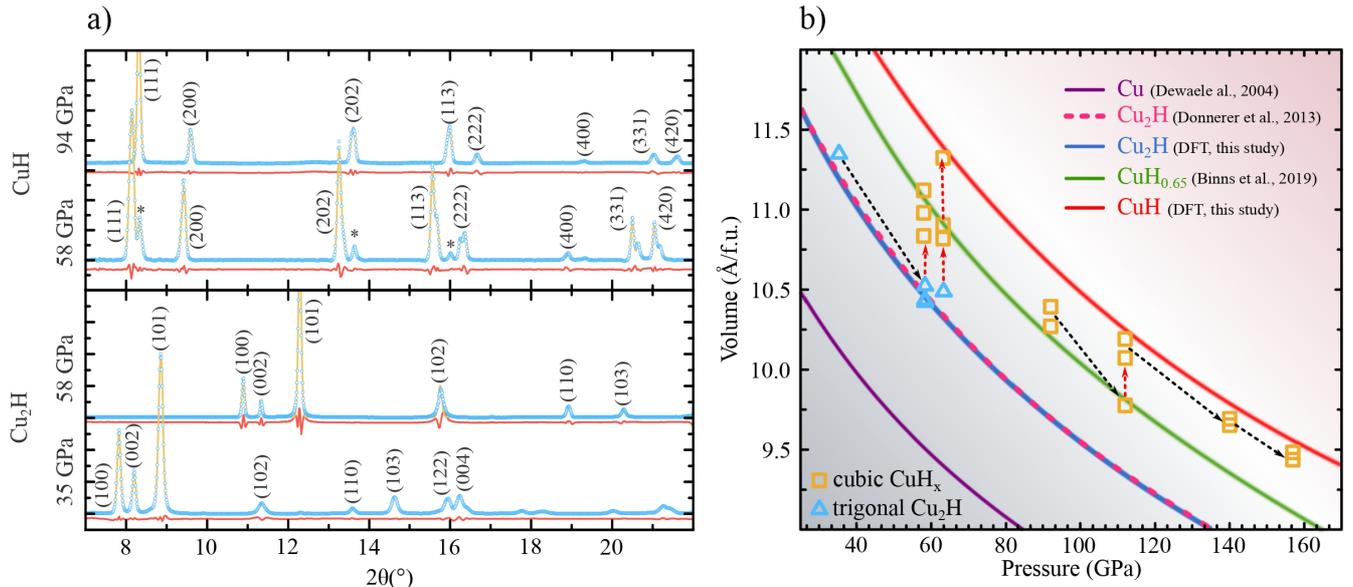

*Figure 1: X-ray diffraction and equations-of-state of Cu and Cu-H phases. a) Examples of powder X-ray diffraction patterns of $Cu_2H$ and CuH synthesized at different pressures ($\lambda=0.41178$ Å; yellow lines are Rietveld fits to the experimental data (blue dots are experimental data points); red lines denote the residuals. Indexed peaks are characteristic to $P\bar{3}m1$ $Cu_2H$ and $Fm\bar{3}m$ CuH structures. Asterisks denote diffraction peaks from micro-diamonds formed after laser heating. b) Equations of state of the copper hydride phases and pure Cu: the equations-of-state for pure copper (Dewaele et al. [21]) and $CuH_{0.65}$ (Binns et al. [18]) are shown in purple and green, respectively. The brown dashed line denotes the equation-of-state for $P\bar{3}m1$ $Cu_2H$ (Donnerer et al. [17]). The results of a third order Birch-Murnaghan fit to the ab-initio energies are shown by a blue curve for $Cu_2H$ and red for CuH. The equation-of-state parameters can be found in Table S2 [20]. Open orange squares and blue triangles denote structures refined from diffraction data, and the arrows illustrate the experimental pathways of heating (red) and compression (black).*

After laser heating at 35 GPa, a trigonal (space group $P\bar{3}m1$) $Cu_2H$ compound formed, as determined with powder (Fig. 1a) and single crystal diffraction (Tables S1 and S3 [20]). This phase remained stable under compression to at least 90 GPa at room $T$. At 35 GPa and 58 GPa, its atomic volumes (per Cu atom, $V$/Cu) are 11.27(2) Å$^3$ and 10.44(1) Å$^3$, in agreement with values expected from the equations-of-state (EOS) previously reported from experiments [17] and computed here (Fig. 1b).

Laser heating of $Cu_2H$ or pristine copper with paraffin at $P>50$ GPa resulted, in agreement with previous observations [18], in the formation of a cubic phase with an fcc arrangement of Cu atoms. The same phase has been found upon laser heating of the samples at pressures up to ~160 GPa, the highest $P$ reached in this study. Its $V$/Cu at 58, 94, and 112 GPa (11.11(2), 10.27(4), 9.80(6) Å$^3$) agree with those derived from EOS of fcc-structured $CuH_{0.65}$ [18] (Fig. 1b).

After repeated heating and $T$-quench of the cubic phase at $P \geq 110$ GPa, an increase of $V$/Cu was observed (Fig. 1b) and values eventually became larger than expected for previously reported copper hydrides: at 157 GPa, for example, $V$/Cu reaches 9.48(2) Å$^3$ vs 9.03 Å$^3$ for $CuH_{0.65}$ [18] and 7.79 Å$^3$ for Cu [21] based on the reported EOS. Thus, our data suggest that at $P \geq 110$ GPa we synthesized a copper hydride with a larger hydrogen content than any previously reported at high $P$ [17,18].

An exact determination of the hydrogen content of Cu-compounds formed in different experiments is difficult, but we note that the DFT-based calculations reproduce the experimental EOS for trigonal $Cu_2H$ well (Fig. 1b). The experimental data compares similarly well with the *ab-initio* calculations on the EOS for NaCl-structured CuH (Fig. 1b), and we therefore infer that the cubic copper hydride we synthesized at high $P$ under $T$-cycling has a ratio of Cu/H=1 (CuH), see Table S4 [20].

Estimating chemical compositions of copper hydrides and assessing their respective stability fields have provided the basis for the discussion and interpretation of $^1$H-NMR data. Figure 2 shows representative $^1$H-NMR spectra of the copper hydrides at different pressures (see Figure S1 for all $^1$H-NMR spectra [20]). The NMR signal from the copper-paraffin sample (5 GPa before heating) originates solely from the paraffin reservoir, and both the full width at half maximum (FWHM) of the NMR signal and the relaxation rate agree well with previous studies for paraffin [10,14]. As the resonance frequency of this signal does not change considerably (Fig. 2), it was used as a reference. Additional signals appeared after laser heating at 50 GPa and 89 GPa. We assigned them to $Cu_2H$ and CuH, respectively. During decompression of $Cu_2H$ to $P\leq16$ GPa, the spectrum was interpreted as that of $CuH_{0.15}$, given the agreement in terms of a transition $P$ with previous experimental work [22]. The successive decrease of resonance shift with increasing amount of hydrogen is in agreement with other metal hydrides synthesized at ambient $P$ [23–27].

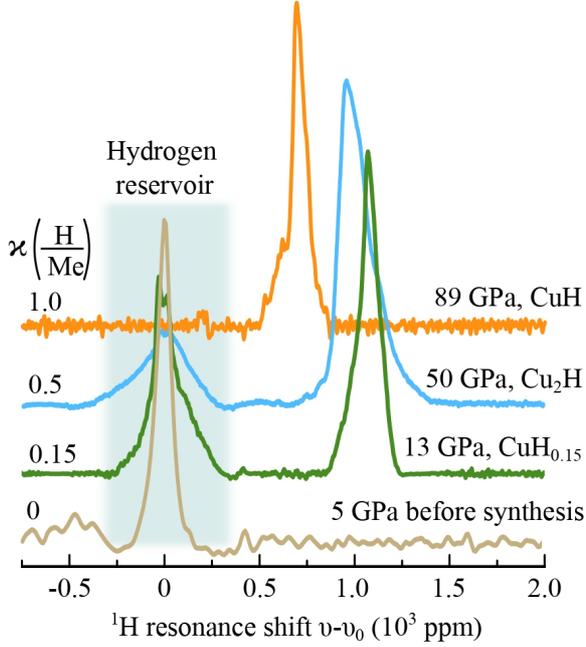

*Figure 2: Representative $^1$H-NMR spectra of copper hydrides, CuH, Cu$_2$H, and CuH$_{0.15}$, formed after laser heating of copper-paraffin samples at 89 GPa and 50 GPa, and on decompression of Cu$_2$H (13 GPa).*

Figure 3a shows the relative changes of the hydrogen Knight shift $K_H$ as a function of compression $V/V_0$ for the three copper hydrides synthesized, CuH, Cu$_2$H, and CuH$_{0.15}$ (for $K_H$ as a function of $P$, see Fig. S2 [20]), as well as for cubic FeH from our previous study [14]. Cu$_2$H and FeH follow a slope expected for a free electron gas-like system ($K_H \propto V^{2/3}$ [14]), while both CuH$_{0.15}$ and CuH violate Fermi gas ideality, indicating that these phases might at best be considered as bad metals or semi metals, in agreement with recent *ab-initio* DFT calculations [28].

The volume dependence of $K_H$ for Cu$_2$H displays a deviation from free electron gas behavior between 43 and 58 GPa, similar to FeH between 64 and 110 GPa as previously described [14]. This effect can also be seen in the dependence of the electronic density of states (DOS) at the Fermi energy $N(E_F)$ from *ab-initio* simulations: They consist of the sum (dashed lines in Fig. 3a) of the 4s electron DOS of Cu and the 1s electron DOS of H. Interestingly, the contribution of the hydrogen 1s electron to $N(E_F)$ for Cu$_2$H and FeH (Fig. 4a) gradually increases with decreasing distance between hydrogens $r_{HH}$ (see also Fig. S3 [20]), similar to the total DOS (Fig. 3a). This observation suggests that conduction electron density from the uncompensated 4s states in Cu$_2$H or respectively the uncompensates 3d-t$_{2g}$ states in FeH [14] is transferred to $N_{H-1s}(E_F)$. By contrast, $N_{H-1s}(E_F)$ of semi-metallic CuH shows a strongly negative slope (Fig. 4a). Given the striking similarity of the observed behavior in $K_H$ and electronic DOS for Cu$_2$H with that of FeH described recently [14], we conclude that they can be explained by the formation of a metallic hydrogen sublattice in trigonal Cu$_2$H at $P \sim 40$ GPa.

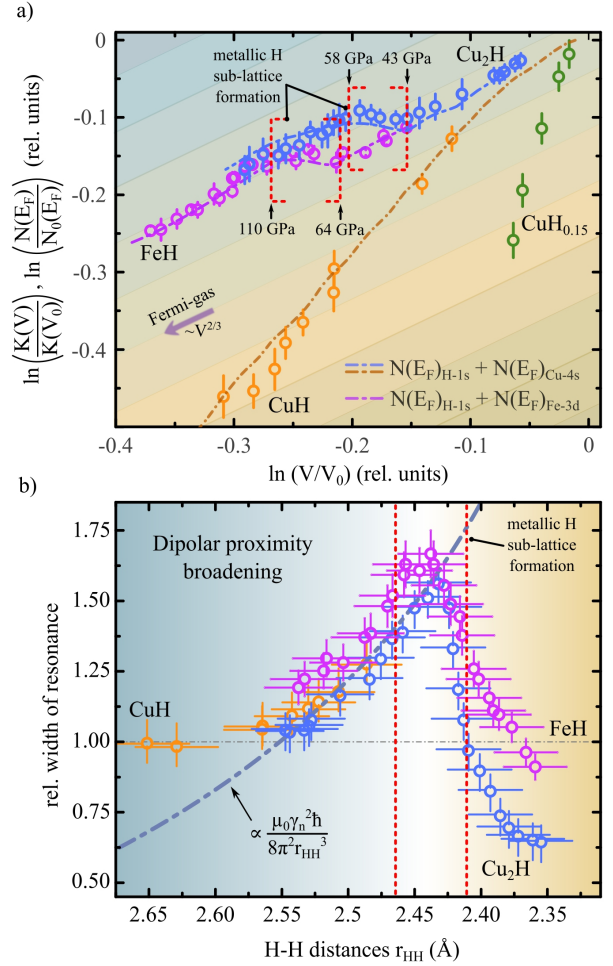

*Figure 3: Comparison of experimental NMR data and electronic structure calculations. a) Double logarithmic power plot of relative changes in the NMR Knight shift $K_H(V)$ and the electron density of states at the Fermi energy $N(E_F)$ as a function of relative volume. Experimental data points (blue, magenta, orange and green) have been normalized to $K_H(V_0)$ and $V_0$ using the respective equation of states from Tkacz et al. [22] (CuH$_{0.15}$) and DFT computations from this study (CuH and Cu$_2$H) (Table S2[20]); The blue, magenta and orange dashed lines (splines through computed values) show the respective volume dependence of $N(E_F)$. Diagonal color stripes are guides to the eye representing a $\propto V^{2/3}$ scaling for free-electron Fermi-gas-like behavior [14]. Black arrows denote respective pressures points. Data and results for FeH were taken from our previous work [14]. b) Relative line-widths of $^1$H-NMR spectra as a function of the H-H distance $r_{HH}$ for CuH, Cu$_2$H, and FeH. The dotted blue line depicts the theoretical line-width dependence for pure dipolar broadening.*

Figure 3b shows the dependence of the relative proton NMR resonance line-widths on $r_{HH}$ (taken from *ab-initio* simulations at corresponding $P$) in both metallic phases of Cu$_2$H and FeH as well as in semi-metallic CuH. For $r_{HH}$>2.45 Å (P=43 GPa for Cu$_2$H and P=64 GPa for FeH), line-widths of the NMR signals for both transition metal hydrides as well as semi-metallic CuH increase with P ($\Delta\nu \propto r_{HH}^{-3}$) due to dipolar couplings of hydrogen spin-1/2 nuclei through a pronounced $\propto r_{HH}^{-3}$ slope [29,30]. At higher P, i.e. $r_{HH}$<2.45 Å, NMR absorption line-widths decrease significantly with P (Fig. S1 in [20]). Since diffraction data and computational results on Cu$_2$H (this study) and FeH [14,19] do not show signs of structural transitions within the experimental P-range, line narrowing effects cannot be related to atomic re-arrangements. Rather, line-widths change for two different transition metal hydrides at comparable $r_{HH}$, suggesting that electronic and dynamic effects on the hydrogen spin system, needs to be considered. Hydrogen spin-lattice relaxation rates $R_1$ (or equivalent relaxation times $T_1 = 1/R_1$, Fig. S2 [20]) in metal hydrides originate from two mechanisms, representing electronic hyperfine interactions $R_1^e$ and proton diffusion $R_1^d$ [31]:

(i) $R_1^e$ of hydrogen nuclei due to interaction with conduction electron spins are described by a Korringa-like behavior for a Fermi-gas [32,33] with

$$K_H^2 = \frac{\hbar}{4\pi k_B T}\left(\frac{\gamma_e}{\gamma_n}\right)^2 R_1^e, \quad (1)$$

where $k_B$ is the Boltzmann and $\hbar$ the reduced Planck's constant, $\gamma_e$ and $\gamma_n$ are the gyromagnetic ratios of the electron and the hydrogen nucleus, respectively. $K_H$ is the hydrogen Knight shift originating from contact hyperfine interaction between the hydrogen 1s electron orbital and the proton nucleus via

$$K_H = \frac{2\mu_0 \mu_B}{A} H_{hf}^s N(E_F), \quad (2)$$

with $\mu_0$ the permeability of free space, A Avogadro's number, $\mu_B$ the Bohr magneton, and the hyperfine field $H_{hf}^s$ [14].

(ii) A modulation of the relaxation rate due to the diffusive motion of hydrogen nuclei ($R_1^d$) in crystalline systems, such as Cu$_2$H and FeH, can be expected to be well described by the theory of Bloembergen, Pound and Purcell [34] by the use of a single correlation time for a stochastic undirected motion of hydrogen atoms as

$$R_1^d = \frac{3\pi}{10}\frac{\gamma_n^4 \hbar^2 N_0}{aD}, \quad (3)$$

where $N_0$ is the number density of of atoms, a is the distance of closest approach which is at the order of $r_{HH}$, h is Planck's constant and D the hydrogen diffusion coefficient.

Calculating the relaxation rate $R_1$ (Fig. S2 [20]) in the extreme narrowing limit [29] ($R_1 = R_1^e + R_1^d$) from the experimental linewidths via the relationship $R_1 \approx \pi \cdot \Delta\nu$ with $\Delta\nu$ the full-width half-maximum line-widths and subtracting the conduction electron contribution $R_1^e$ calculated via eq. (1) – following previous work at ambient P [35,36] – gives access to D of proton in transition metal hydrides at high P in the DAC. Figure 4b shows experimentally derived and computed D as a function of $r_{HH}$ for Cu$_2$H and FeH, and semi-metallic CuH. Experimental estimates for D range from ~3.8(4) ·10$^{-7}$ cm$^2$/s and ~3.3(6) ·10$^{-7}$ cm$^2$/s at 0 GPa to ~1.6(9) ·10$^{-7}$ cm$^2$/s and ~0.8(9) ·10$^{-7}$ cm$^2$/s at highest P for Cu$_2$H and FeH, respectively. The steeper slope of D at $r_{HH}$ < 2.42 Å coincides with a decrease in spin-lattice relaxation time $T_1$ starting at a compression of $V/V_0 \approx 0.8$, observed in both Cu$_2$H (Fig. S2 [20]) and FeH. Proton diffusivity in CuH was found approximately two orders of magnitude smaller than for the Cu$_2$H phase. Hydrogen diffusion coefficients calculated by *ab-initio* molecular dynamic (Figs. 4b and S4 [20]) agree with our $^1$H-NMR derived results.

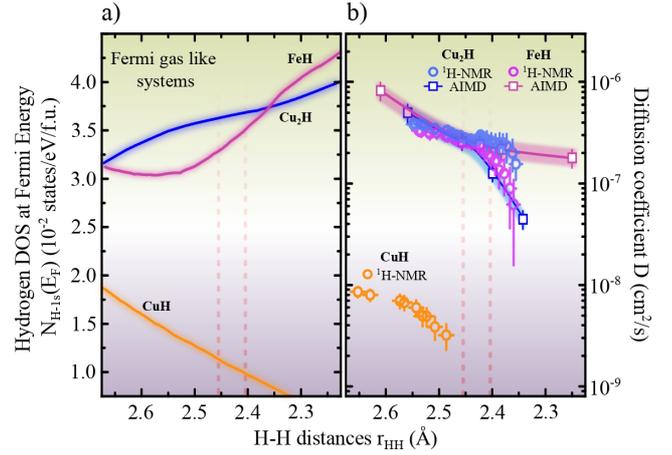

*Figure 4: Comparison of electronic structure and proton diffusivities. a) Hydrogen 1-s electron density of state contribution $N_{H-1s}(E_F)$ of CuH, Cu$_2$H and FeH as a function of the H-H distance in the structures (pressure increasing to the right). b) Protondiffusion coefficients extracted from NMR data (blue, purple and orange circles) and computed via ab-initio molecular dynamics simulations (blue and magenta squares and lines) for metallic Cu$_2$H and FeH, and the semimetal CuH. Vertical red lines denote the inferred metallic hydrogen sub-lattice formation range.*

Hydrogen mobility investigated in other metal hydride systems by means of standard as well as pulsed-field gradient NMR found self-diffusion coefficients to typically lie between 10$^{-13}$ and 10$^{-7}$ cm$^2$/s for a broad range of compounds [31,37–39]. Diffusivities for Cu$_2$H and FeH fall in the upper limit of this broad range, suggesting that proton self-diffusion plays a significant role in high-P metal hydrides compared to hydrides stabilized at ambient conditions. Given the fact that comparable metal hydrides, such as TiH$_{1.66}$ [39] and VH$_{0.39-0.85}$ [35], show significantly lower proton diffusivities at the order of 10$^-$ $10^{-13}$ to $10^{-10}$ cm$^2$/s at larger $r_{HH}$, it stands to reason that proton self-diffusion in H-rich super-hydrides, such as LaH$_{10}$, YH$_{10}$ or CaH$_6$ with $r_{HH} < 1.2$ Å [40–42], approaches values several orders of magnitude larger than for Cu$_2$H and FeH.

The presented work on Cu$_2$H reveals a surprising onset of hydrogen-hydrogen interactions in a *P*-range 43-58 GPa, associated with the formation of a metallic hydrogen sublattice characteristic for novel metals hydrides [40]. Strikingly, our previous work on FeH shows very similar characteristic effects in the electronic structure in a *P*-range of 64-110 GPa. Correlating these effects with the average hydrogen-hydrogen distances, $r_{HH}$, we find that in both compounds metallic hydrogen sub-lattices form at almost identical hydrogen distances, implying that the effect may only weakly depend on the parent transition metal ions. This insight may provide an important step in the future search and design of novel hydride-based high-*T* superconductors.

**Acknowledgements:** We thank Nobuyoshi Miyajima for help with the FIB milling. The authors thank the German Research Foundation (Deutsche Forschungsgemeinschaft, DFG, Project Nos. DU 954/11-1, DU 393/13-1, DU 393/9-2, ME 5206/3-1, and STE 1105/13-1 in the research unit FOR 2440) and the Federal Ministry of Education and Research, Germany (BMBF, Grant No. 05K19WC1) for financial support. D.L. thanks the Alexander von Humboldt Foundation for financial support. N.D. thanks the Swedish Government Strategic Research Area in Materials Science on Functional Materials at Linköping University (Faculty Grant SFO-Mat-LiU No. 2009 00971).

**Author Contributions:**
T.M. and F.T. contributed equally to this study.
T.M. and L.D. designed the experiments. T.M., L.D., S.K. and D.L. prepared all diamond anvil cells. G.C., D.L., E.K., T.F., K.G., M.H. and L.D. performed the diffraction, T.M. and G.C. the NMR experiments. F.T. and G.S.N. designed and conducted the *ab-initio* calculations. T.M. and F.T. analyzed NMR and XRD data and *ab-initio* results. T.M., G.S.N., F.T., L.D., and N.D. assessed all data and wrote the manuscript.

# Metallic Hydrogen Sublattice and Proton Mobility in Copper Hydride at High Pressure
# -
# Supplemental Material

Thomas Meier[1]*, Florian Trybel[1], Giacomo Criniti[1], Dominique Laniel[2], Saiana Khandarkhaeva[1], Egor Koemets[1], Timofey Fedotenko[2], Konstantin Glazyrin[3], Michael Hanfland[4], Gerd Steinle-Neumann[1], Natalia Dubrovinskaia[2,5] and Leonid Dubrovinsky[1]

[1]Bayerisches Geoinstitut, Universität Bayreuth, 95440 Bayreuth, Germany
[2]Material Physics and Technology at Extreme Conditions, Laboratory of Crystallography, Universität Bayreuth, 95440 Bayreuth, Germany
[3]Deutsches Elektronen-Synchrotron (DESY), 22603, Hamburg, Germany
[4]European Synchrotron Radiation Facility (ESRF), 38043 Grenoble Cedex, France
[5]Department of Physics, Chemistry and Biology (IFM), Linköping University, SE-581 83, Linköping, Sweden
*thomas.meier@uni-bayreuth.de

## Experimental Details

### NMR studies

Different sets of diamond anvil cells (DAC) were prepared for nuclear magnetic resonance (NMR) and X-ray diffraction (XRD) studies. Gaskets were manufactured by pre-indenting a 250 μm thick rhenium foil to about 25 and 15 μm, respectively, and the hole for the sample chamber was laser drilled in the center of the flat indentation. All cells were loaded with paraffin oil (Sigma Aldrich Ltd.), serving as a hydrogen reservoir, NMR reference signal and a pressure-transmitting medium. High quality copper powder (5N purity) was added. The volume ratio of paraffin oil to copper powder was estimated to be 1:10 to 1:15, in order to ensure hydrogen excess. Sample synthesis by reacting copper with paraffin was conducted at varying DAC pressures ($P$) by first compressing the sample cell and then using double sided laser heating in continuous wave mode with a nominal laser power of about 25-40 W to create temperatures ($T$) in excess of 2000 K. Pressure was monitored using the first order Raman spectrum of the diamond edge [1,2].

Four DACs for NMR measurements were prepared. Diamond anvils with pairs of 250 μm and 100 μm culets were used. Anvils were covered with a 1 μm layer of copper using physical vapor deposition. By focused ion beam milling, Lenz lens structures were cut from the copper layer to form a double- and triple stage NMR resonator setup for the 250 μm and 100 μm anvils, respectively. To insure insulation between the resonators and the metallic gasket, the gasket was covered by 500 nm of Al$_2$O$_3$ using chemical vapor deposition. Helmholtz excitation coils were made from 80 μm thick PTFE (Teflon) insulated copper wire consisting of 4 to 5 turns with an inner diameter of about 3 mm. Both coils were fixed onto the diamond backing plates such that the copper coated anvils are at the coils' center. After cell loading and closure, the excitation coils were connected to form a Helmholtz coil-like arrangement. The NMR cells were integrated into a standard in-house built NMR probe for use at a magnetic field of about 1030 mT generated by an electron spin resonance electromagnet, corresponding to a resonance frequency of the hydrogen nuclei of 43,851 MHz. In order to limit radio frequency excitation to a region of interest of about 2000 ppm (90 kHz at 1030 mT), we recorded high-$P$ $^1$H-NMR spectra using amplitude-modulated band-selective pulses with uniform response and pure phase (E-BURP) as described by Geen and Freeman [3]. $T_1$ relaxation times were obtained using an I-BURP/E-BURP inversion recovery.

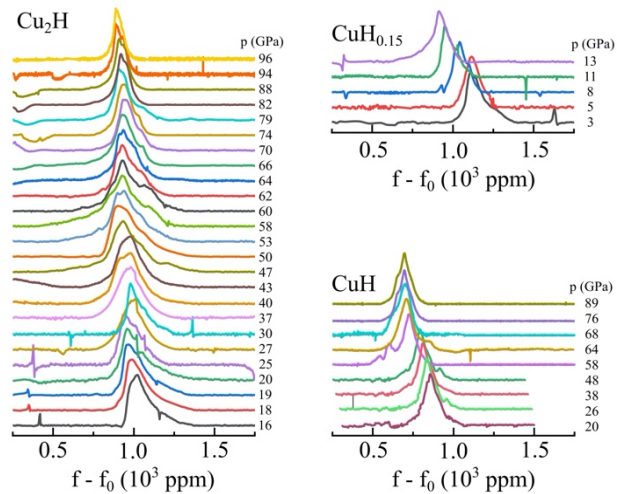

Figure S1: $^1$H-NMR spectra of copper hydrides synthesized in this study.

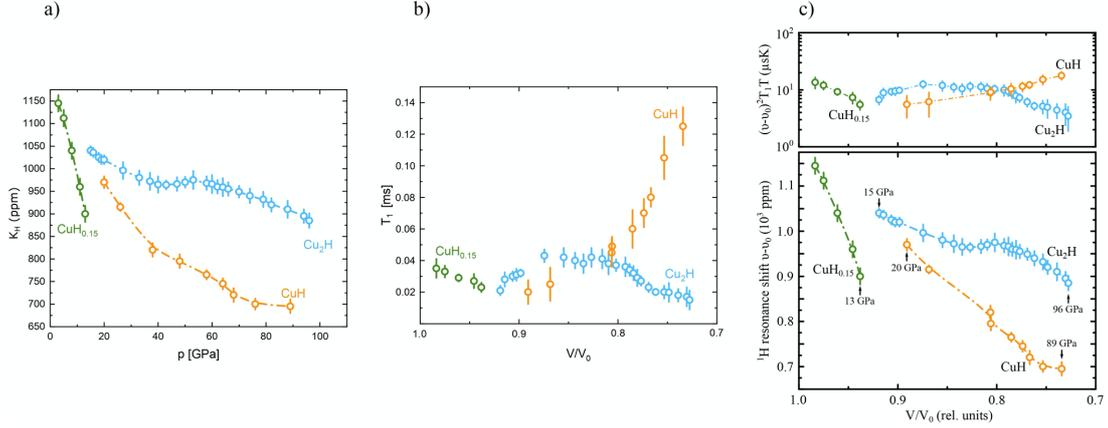

*Figure S2: a) Knight shift $K_H$ versus pressure P, b) spin lattice relaxation times $T_1$ of the three copper hydrides synthesized and c) summary of Knight shift data (bottom) as well as Korringa ratios (top) as a function of compression $V/V_0$ using $V_0$ from the equations of state given in Table S2).*

**X-ray diffraction data**

Two DACs with culets of 250 μm and 120 μm in diameter were used for XRD experiments. Temperature was determined by multiwavelength spectroradiometry. Reaction of paraffin oil with copper was confirmed by a change in optical properties of the sample and by the appearance of the Raman peak of micro-diamonds formed by the carbon in the paraffin oil. Pressure inside the DAC was determined using the equations of state of unreacted copper (Cu), rhenium gasket material (Re) and microdiamonds (Dia) produced during laser heating by reaction of Cu and paraffin oil; $P$ estimated from the Raman shift of micro-diamonds [4] was consistent with that obtained from the diamond edge of the anvils used for $P$-determination in the NMR experiments. High-resolution XRD measurements were performed at beamlines ID15 of the European Synchrotron Radiation Facility (ESRF) in Grenoble, France, and P02.2 of Petra III (DESY) in Hamburg, Germany. At beamline ID15, an X-ray beam of 30.142 keV (0.41134 Å) and a large-area MAR555 flat panel detector were used; wide-scan images were collected upon continuous rotation of the cell between –20° to + 20° ω. At beamline P02.2, an X-ray beam of 42.87 keV (0.2982 Å) and a Perkin Elmer XRD1621 flat panel detector were used; wide-scan images were collected upon continuous rotation of the cell between –20° to + 20° ω and still images were collected *in-situ* during laser heating at 112 GPa and 2000 K.

Phase identification was carried out from powder diffraction profiles integrated from still and wide-scan images, using the DIOPTAS software [5]. Cell parameters were obtained from LeBail refinement of selected powder diffraction patterns using JANA2006 [6]. A background correction was applied either manually or by using a Chebyshev polynomial, and peaks were fitted with pseudo-Voigt functions. $Cu_2H$ was confirmed to have a distorted *hcp* structure with trigonal symmetry, based on the presence of *hcp*-forbidden peaks (e.g. 001, 003). CuH exhibits an *fcc* NaCl-type structure, but shows an increase in volume that ranges from approx. 16% to 21% relative to pure copper in good agreement with our ab-initio calculations. Laser heating at 140 GPa produced virtually no further increase in the volume of CuH, implying that either the H reservoir was exhausted or that the Cu hydride reached its maximum H storing capacity.

The analysis of single-crystal data, including integration of intensities, frame scaling and empirical absorption corrections, was carried out using *CrysAlis^Pro*. Structural solution and refinement were performed using the ShelX package [7,8] in the WinGX software [9].

*Table S1: Summary of experimental data from diffraction experiments. Pressure (P) inside the DAC was determined using the equations of state (EOS) of unreacted copper (Cu), rhenium gasket material (Re) and microdiamonds (Dia) produced during laser heating by reaction of Cu and paraffin oil. XRD measurements were performed before (Cold) and after heating, and the temperature (T) at which each phase was synthesized was measured through multiwavelength spectroradiometry. Copper hydride phases identified and analyzed at each pressure point are reported as $Cu_2H$ (trigonal phase), $CuH_x$ (cubic phase, with $0.5 < x < 1$) and CuH (cubic phase, full H occupancy). Their atomic volume per a Cu atom (V/Cu) reported in Å$^3$, are listed in the last column.*

| Cell | P (GPa) | EOS | T (K) | Phase | V/Cu (Å$^3$) |
|---|---|---|---|---|---|
| GC001 | 35(1) | Cu | 1900(100) | $Cu_2H$ | 11.27(2) |
|  | 35(1) | Cu | 1900(100) | $Cu_2H$ | 11.35(2) |
|  | 58(1) | Cu, Dia | Cold | $Cu_2H$ | 10.53(4) |
|  | 58(1) | Cu, Dia | Cold | $Cu_2H$ | 10.42(2) |
|  | 58(1) | Cu, Dia | Cold | $Cu_2H$ | 10.44(2) |
|  | 58(1) | Cu, Dia | 2000(100) | $CuH_x$ | 10.83(2) |
|  | 58(1) | Cu, Dia | 2000(100) | $CuH_x$ | 10.91(3) |
|  | 58(1) | Cu, Dia | 2000(100) | $CuH_x$ | 11.11(2) |
| GC003 | 94(2) | Dia | Cold | $CuH_x$ | 10.21(3) |
|  | 94(2) | Dia | 2000(100) | $CuH_x$ | 10.30(6) |
|  | 94(2) | Dia | 2000(100) | $CuH_x$ | 10.27(4) |
|  | 94(2) | Dia | 2000(100) | $CuH_x$ | 10.28(3) |
|  | 112(2) | Cu, Re | Cold | $CuH_x$ | 9.80(6) |
|  | 112(2) | Cu, Re | 2000(300) | CuH | 10.08(2) |
|  | 112(2) | Cu, Re | 2000(300) | CuH | 10.08(3) |
|  | 140(2) | Dia | 1800(100) | CuH | 9.66(2) |
|  | 140(2) | Dia | 1800(100) | CuH | 9.67(3) |
|  | 157(3) | Dia | 2000(100) | CuH | 9.48(2) |
|  | 157(3) | Dia | 2000(100) | CuH | 9.44(3) |
|  | 157(3) | Dia | 2000(100) | CuH | 9.48(2) |
| TM003 | 63(1) | Dia | Cold | $Cu_2H$ | 10.49(6) |
|  | 63(1) | Dia | Cold | $Cu_2H$ | 10.49(4) |
|  | 63(1) | Dia | 1800(200) | $CuH_x$ | 11.33(6) |
|  | 63(1) | Dia | 1800(200) | CuH | 10.90(3) |
|  | 63(1) | Dia | 1800(200) | CuH | 10.82(4) |

**Equations of state**

*Table S2: Equation of state parameter for pure copper, the three copper hydrides and fcc iron hydride used in the conversion of volumes to pressure in this study. $V_0$ is the equilibrium volume, $K_0$ the bulk modulus at zero pressure, and $K_0'$ its pressure derivative.*

|  | Reference | Space Group | $V_0$ p.f.u (Å$^3$) | $K_0$ (GPa) | $K_0'$ |
|---|---|---|---|---|---|
| Cu | [10] | $Fm\bar{3}m$ | 11.8 | 135 | 4.9 |
| CuH | DFT, This study | $Fm\bar{3}m$ | 14.8 | 132 | 4.5 |
| $Cu_2H$ | [11] | $P\bar{3}m1$ | 13.8 | 138 | 4.6 |
| $Cu_2H$ | DFT, This study | $P\bar{3}m1$ | 13.3 | 138 | 4.6 |
| $CuH_{0.65}$ | [12] | $Fm\bar{3}m$ | 14.1 | 145 | 4.0 |
| FeH | [13] | $Fm\bar{3}m$ | 13.5 | 99 | 11.7 |

Table S3: Refinement and crystal structure details of Cu$_2$H at 35 GPa based on single crystal/powder X-ray diffraction data.

| Chemical formula | Cu$_2$H |
|---|---|
| **Crystal data** | |
| $M_r$ | 64.05 |
| Crystal system, space group | Trigonal, $P\bar{3}m1$ |
| Pressure (GPa) | 35 |
| $a, c$ (Å) | 2.5006(6), 4.1635(15) |
| $V$ (Å$^3$) | 22.546(14) |
| $Z$ | 1 |
| Radiation type | Synchrotron, λ = 0.41134 Å |
| μ (mm$^{-1}$) | 45.95 |
| **Data collection** | |
| Diffractometer | ID15B@ESRF |
| Absorption correction | Multi-scan *CrysAlis PRO* 1.171.38.46 (Rigaku Oxford Diffraction, 2018) Empirical absorption correction using spherical harmonics, implemented in SCALE3 ABSPACK scaling algorithm. |
| No. of measured, independent and observed [$I > 2\sigma(I)$] reflections | 61, 32, 32 |
| $R_{int}$ | 0.014 |
| $(\sin \theta/\lambda)_{max}$ (Å$^{-1}$) | 0.435 |
| **Refinement** | |
| $R[F^2 > 2\sigma(F^2)]$, $wR(F^2)$, $S$ | 0.089, 0.227, 1.56 |
| No. of reflections | 32 |
| No. of parameters | 5 |
| Δρ$_{max}$, Δρ$_{min}$ (e Å$^{-3}$) | 5.94, −3.02 |
| **Atomic coordinates** | |
| Cu1 | (0.6667 0.3333 0.2503(17)) |
| H1 (implied) | (0 0 0) |

*Table S4: Refinement and crystal structure details of CuH at 112 GPa based on single crystal/powder X-ray diffraction data.*

| Chemical formula | CuH |
|---|---|
| **Crystal data** | |
| $M_r$ | 64.55 |
| Crystal system, space group | Cubic, $F\mathrm{m}\bar{3}\mathrm{m}$ |
| Pressure (GPa) | 112 |
| $a$ (Å) | 3.4227(10) |
| $V$ (Å$^3$) | 40.10(3) |
| $Z$ | 4 |
| Radiation type | Synchrotron, λ = 0.2982Å |
| μ (mm$^{-1}$) | 51.68 |
| **Data collection** | |
| Diffractometer | P02.2@PETRA III |
| Absorption correction | Multi-scan *CrysAlis PRO* 1.171.38.46 (Rigaku Oxford Diffraction, 2018) Empirical absorption correction using spherical harmonics, implemented in SCALE3 ABSPACK scaling algorithm. |
| No. of measured, independent and observed [$I > 2\sigma(I)$] reflections | 56, 19, 17 |
| $R_{int}$ | 0.068 |
| $(\sin \theta/\lambda)_{max}$ (Å$^{-1}$) | 0.528 |
| **Refinement** | |
| $R[F^2 > 2\sigma(F^2)]$, $wR(F^2)$, $S$ | 0.070, 0.185, 1.43 |
| No. of reflections | 19 |
| No. of parameters | 2 |
| $\Delta\rho_{max}$, $\Delta\rho_{min}$ (e Å$^{-3}$) | 2.68, -4.07 |
| **Atomic coordinates** | |
| Cu1 | (0 0 0) |
| H1 (implied) | (0.5 0 0) |

## Ab-initio calculations

All *ab-initio* calculations were performed with Quantum Espresso [14,15] with the projector augmented wave (PAW) approach [16]. We used the generalized gradient approximation by Perdew-Burke-Enzerhofer (PBE) [17] to exchange and correlation with the corresponding PAW-PBE potential files. For both Fe and Cu a valence electron configuration that includes electronic states 3s and higher is applied, appropriate to extreme pressure [18]. Convergence tests for the static electronic structure simulations led to a cutoff energy for the plane-wave expansion of 170 Ry (Cu) and 120 Ry (Fe) and to reciprocal space sampling with a Monkhorst-Pack [19] grid of 32x32x32. In addition to the total electronic density of states, site and orbital projections were computed (Fig. S3), with particular attention to the crystal field splitting of the 3d electronic states. Equation of state for CuH and $CuH_2$ were fit on the ground state energies for various unit cell volumes using a Birch-Murnaghan formulation [20]. Fit parameters are summarized in Table S2.

Molecular dynamics (MD) simulations were performed for cells of 54 and 96 atoms for $Cu_2H$ and FeH, respectively, with a Monkhorst-Pack [19] grid of 2x2x2. Calculations were performed in the canonical ensemble, and *T* is controlled by the Anderson thermostat [21]. We employed a timestep of 0.1 fs to account for the fast hydrogen dynamics [22] and run the simulations for a duration of 1.6 ps. Diffusion coefficients of H in these structures were calculated from the mean square displacements obtained via MD trajectories (Fig. S4).

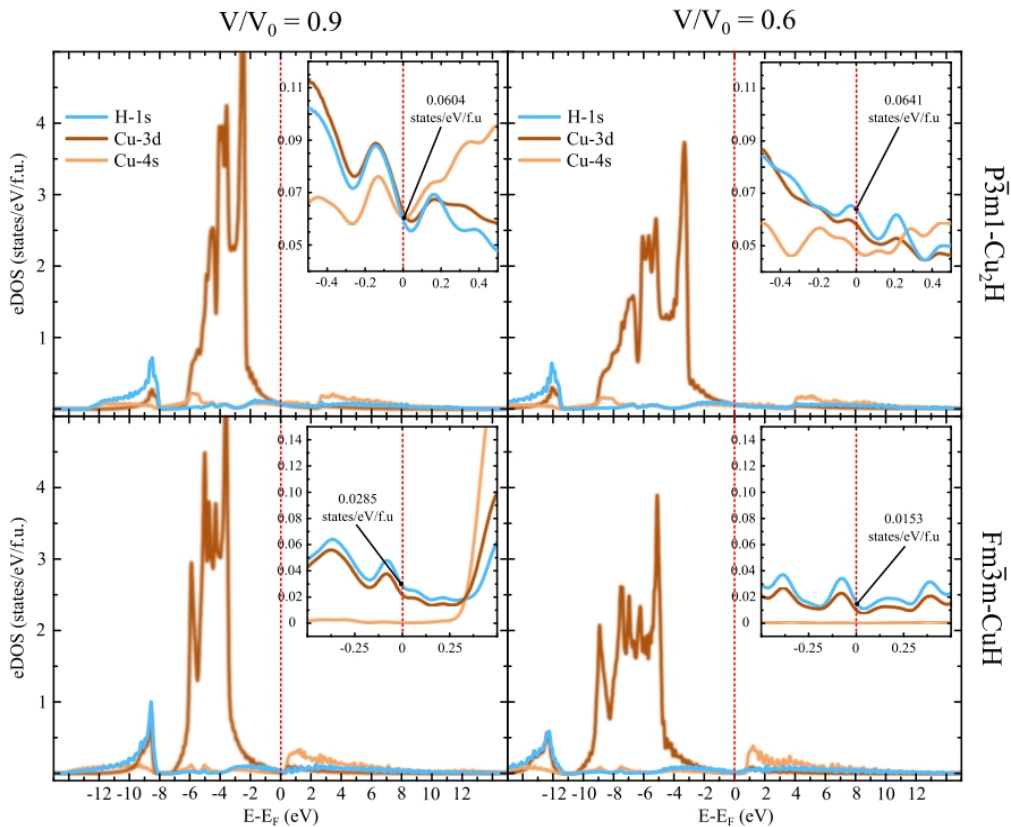

*Figure S3: Partial electron density of states from* ab-initio *DFT calculations for both $Cu_2H$ (top) and CuH (bottom) at compressions of $V/V_0=0.9$ )(left) and 0.6 (right). Insets show close-ups of the vicinity around the Fermi energy $E_F$.*

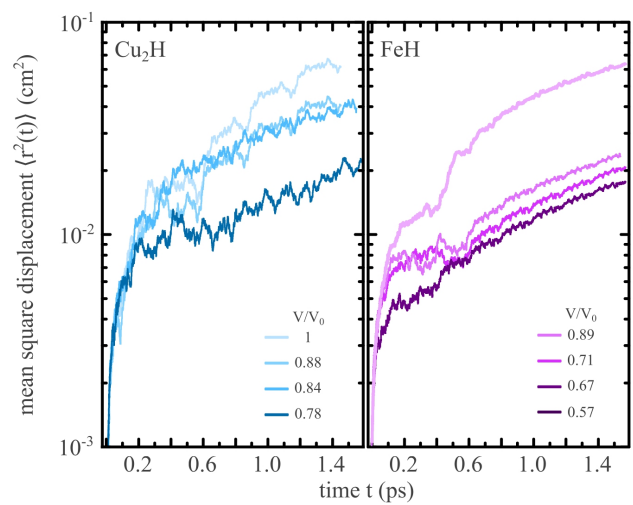

*Figure S4: Mean square displacement of hydrogen atoms as function of time as obtained by means of ab-initio molecular dynamics simulations of $Cu_2H$ and FeH at different compressions.*